\documentstyle[multicol,aps,prl,epsfig]{revtex}

\renewcommand{\narrowtext}{\begin{multicols}{2} \global\columnwidth20.5pc}
\renewcommand{\widetext}{\end{multicols} \global\columnwidth42.5pc}

\input{epsf.tex}

\def\top#1{\vskip #1\begin{picture}(290,80)(80,500)\thinlines \put(
65,500){\line( 1, 0){255}}\put(320,500){\line( 0, 1){
5}}\end{picture}}
\def\bottom#1{\vskip #1\begin{picture}(290,80)(80,500)\thinlines \put(
330,500){\line( 1, 0){255}}\put(330,500){\line( 0, -1){
5}}\end{picture}}

\begin{document}

\draft


\title{Exactly Soluble Model for Umklapp Scattering at
Quantum-Hall Edges}

\author{U.~Z\"ulicke}

\address{Department of Physics, Indiana University, Bloomington,
Indiana 47405 \\ and Institut f\"ur Theoretische
Festk\"orperphysik, Universit\"at Karlsruhe, D-76128 Karlsruhe,
Germany\cite{preadd}}

\date{submitted to Phys. Rev. Lett. on 19 July 1999}

\maketitle

\begin{abstract}

We consider the low-energy, long-wave-length excitations of a
reconstructed quantum-Hall edge where three branches of chiral
one-dimensional edge excitations exist. We find that, in
addition to forward scattering between the three edge-excitation
branches, Coulomb interaction gives rise to an Umklapp-type
scattering process that cannot be accounted for within a
generalized Tomonaga-Luttinger model. We solve the theory
including Umklapp processes exactly in the long-wave-length limit
and calculate electronic correlation functions.

\end{abstract}


\narrowtext

Two-dimensional (2D) electron systems subject to
perpendicular magnetic fields have low-lying excitations that
are localized at the sample boundary~\cite{volkmikh}. At certain
values of the filling factor $\nu$ when the quantum Hall (QH)
effect occurs, the bulk of the 2D system turns out to be
incompressible~\cite{ahmintro}, and the edge excitations comprise
the {\em only} low-lying excitations present.

The detailed electronic structure at the edge of QH systems
depends sensitively on the interplay between the external
potential confining the electrons to the finite sample,
electrostatic repulsion, as well as exchange and correlation
effects. For an ultimately sharp~\cite{sharpness} edge, a single
branch of chiral one-dimensional (1D) excitations is predicted to
exist when the filling factor $\nu=1/m$ where $m$ is an odd
integer~\cite{ahm:wen:90}. In that case, the dynamics of edge
excitations can be described~\cite{wen:rev} using a
Tomonaga-Luttinger (TL) model~\cite{tom:lutt} with only the
right-moving~\cite{rightmove} degrees of freedom present.
However, for a confining potential that is just not sharp enough
to stabilize a single-branch edge, a different configuration is
realized where a lump of electron charge is separated from the
bulk of the QH sample~\cite{ahm:aust:93,wen:prb:94}. Such a
{\em reconstructed\/} edge~\cite{otherrec,exprec,rennapology}
supports three branches of chiral 1D edge excitations, two
right-moving and one left-moving. For even weaker confining
potential, further reconstructions occur, leading to a
proliferation of edge-excitation modes~\cite{manymodes}.
The microscopic structure of a very smooth edge is dominantly
determined by electrostatics, which favors a phase separation of
the 2D electron system at the edge into a series of alternating
compressible and incompressible strips~\cite{smooth}.

Effective TL theories describing single-branch and multi-branch
QH edges predict Luttinger-liquid behavior~\cite{wen:rev}, i.e.,
power laws governing the energy dependence of electronic
correlation functions. The characteristic exponents of these
power laws depend, in general, on details of the microscopic edge
structure. However, in the absence of coupling between different
chiral edge branches or, in some cases, due to disorder
effects~\cite{mpaf:prb:95a}, power-law exponents turn out to be
universally dependent on the bulk filling factor. At present,
microscopic details of the edge structure that is realized in
experiment~\cite{amc:prl} are not fully known. To facilitate a
realistic comparison between theory and experiment, it is
necessary to study the low-lying edge excitations of
reconstructed and smooth edges and investigate the effect
interactions have on the Luttinger-liquid power-law exponents
when more than one branch of edge excitations is present. Most
importantly, it turns out that edges having at least three chiral
branches of 1D edge excitations can support a new kind of
scattering process which does not conserve particle number in
each branch separately. This new interaction process cannot be
accounted for within an effective TL-model description. Here we
study an exactly soluble model for the new scattering process and
determine its effect on the Luttinger-liquid behavior of QH
edges.

\begin{figure}
\centerline{\epsfig{file=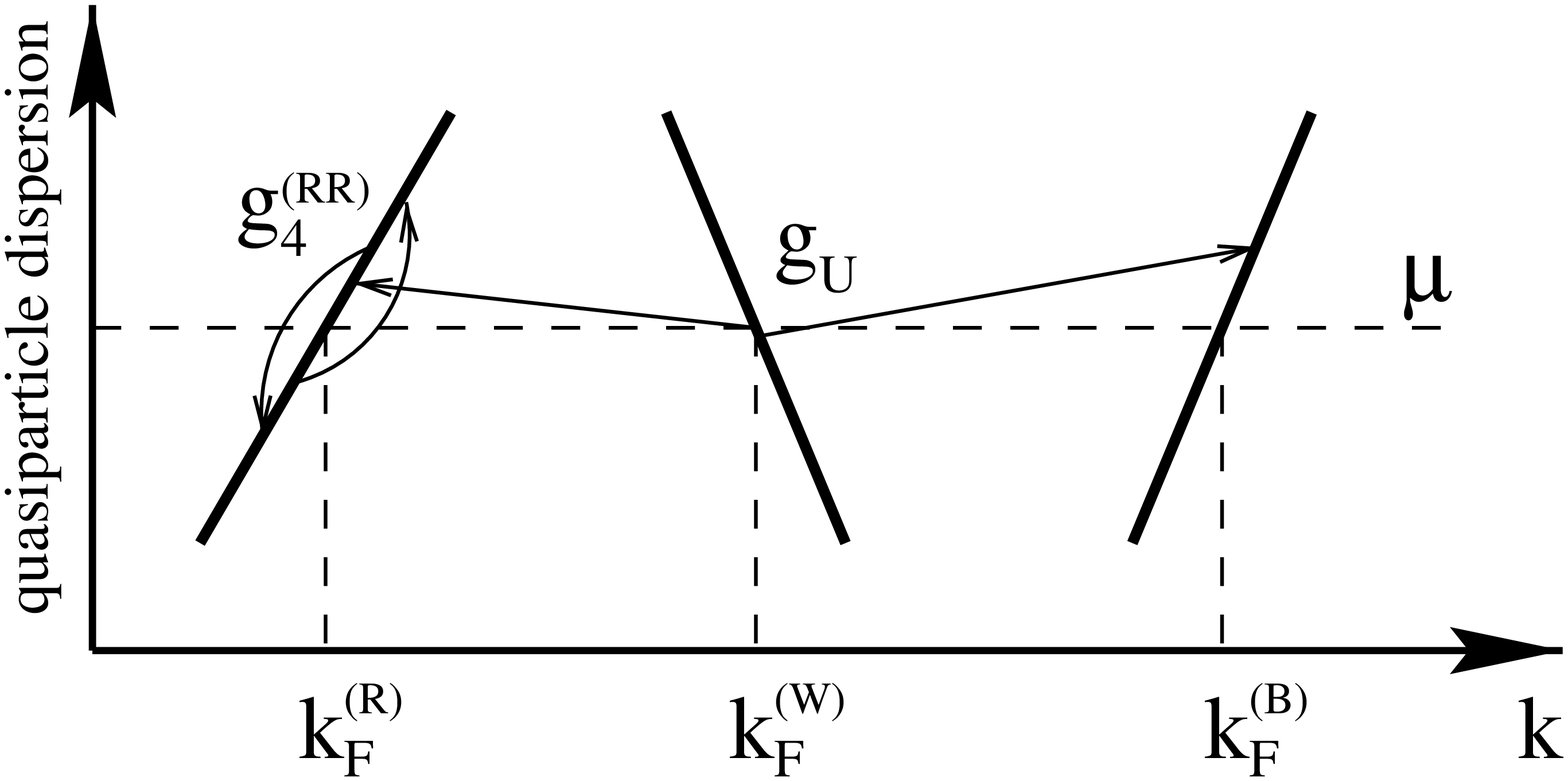,width=3in}}
\caption{Schematic depiction of quasiparticle dispersion and 
interaction processes at a reconstructed QH edge. At low
energies, only interaction processes involving electrons close to
the Fermi points $k_{\text{F}}^{\text{(R)}}$,
$k_{\text{F}}^{\text{(W)}}$, and $k_{\text{F}}^{\text{(B)}}$ are
important. As examples, we show forward scattering within the
R-branch (coupling constant $g_4^{\text{(RR)}}$) and the Umklapp
process ($g_{\text{U}}$).}
\label{scattfig}
\end{figure}

We focus on the edge of a QH sample at $\nu=1$ that has undergone
reconstruction such that three branches of edge excitations are
present. To be specific, we choose the Landau gauge where
lowest-Landau-level (LLL) basis states $\chi_k(x,y) = \Phi_k(y)\,
\exp\{ikx\}/\sqrt{L}$ are labeled by a 1D wave vector $k$. Here,
$\ell=\sqrt{\hbar c/|e B|}$ denotes the magnetic length, $L$ is
the edge perimeter, and $\Phi_k(y) = \exp\{-(y - k\ell^2)^2/(2
\ell^2)\}/\sqrt{\pi^{1/2}\ell}$. In the absence of interactions
between different edge branches, the ground state will be a
generalized Fermi-sea state that is a Slater determinant of LLL
basis states whose wave-vector label satisfies
$k\le k_{\text{F}}^{\text{(R)}}$ or $k_{\text{F}}^{\text{(W)}}\le
k\le k_{\text{F}}^{\text{(B)}}$. The Fermi `surface' consists of 
three (Fermi) points $k_{\text{F}}^{\text{(R)}} < 
k_{\text{F}}^{\text{(W)}}<k_{\text{F}}^{\text{(B)}}$. As in
Tomonaga's approach to interacting 1D electron
systems~\cite{tom:lutt}, long-wave-length electronic excitations
at the reconstructed edge can be identified according to which
Fermi point they belong to. This makes it possible to rewrite the
long-wave-length part of the electron operator as follows:
\widetext
\top{-2.8cm}
\begin{equation}
\psi(x, y) = \Phi_{k^{\text{(R)}}_{\text{F}}}(y)\,
e^{i k^{\text{(R)}}_{\text{F}} x}\, \psi^{\text{(R)}}(x) +
\Phi_{k^{\text{(W)}}_{\text{F}}}(y)\, e^{i
k^{\text{(W)}}_{\text{F}} x}\, \psi^{\text{(W)}}(x) +
\Phi_{k^{\text{(B)}}_{\text{F}}}(y)\, e^{i
k^{\text{(B)}}_{\text{F}} x}\, \psi^{\text{(B)}}(x)\quad .
\end{equation}
\narrowtext
\noindent
The operator $\psi^{\text{(R,W,B)}}(x)$ creates an electron
belonging to the chiral 1D edge branch labeled R, W, B,
respectively. The original 2D interacting Hamiltonian
specializes, in the low-energy limit, to an effective 1D
Hamiltonian involving scattering processes of electrons close to
the three Fermi points. We can classify these interaction
processes in terms of a generalized $g$-ology scheme that is
familiar from the treatment of conventional interacting 1D
electron systems~\cite{sol:adv:79}. (See Fig.~\ref{scattfig}.)
Forward scattering and backscattering~\cite{backscatt} can be
accounted for within a generalized TL model~\cite{wen:prb:94}:
\begin{equation}\label{TLmodel}
H_{\text{TL}} = \frac{1}{2 L} \sum_{q\ne 0} \big[\vec \varrho_q
\big]^\dagger \left(2\pi\hbar {\mathcal K} + {\mathcal V}_q
\right)\vec \varrho_{-q} \quad .
\end{equation}
Here we defined the vector $\vec \varrho_{q} =
\big(\varrho_q^{\text{(B)}}, \varrho_q^{\text{(R)}},
\varrho_q^{\text{(W)}}\big)$ of density fluctuations at the three
Fermi points, and matrices ${\mathcal K} = {\mathrm{diag}}\big(
v_{\text{F}}^{\text{(B)}}, v_{\text{F}}^{\text{(R)}},
v_{\text{F}}^{\text{(W)}}\big)$ and
\begin{equation}
{\cal V}_q  = \left( \begin{array}{ccc}
g_4^{\text{(BB)}}(q) & g_4^{\text{(BR)}}(q) &
g_2^{\text{(BW)}}(q) \\ g_4^{\text{(BR)}}(q) &
g_4^{\text{(RR)}}(q) & g_2^{\text{(RW)}}(q) \\
g_2^{\text{(BW)}}(q) & g_2^{\text{(RW)}}(q) &
g_4^{\text{(WW)}}(q) \end{array} \right) \quad .
\end{equation}
It is straightforward to diagonalize $H_{\text{TL}}$ by a
transformation $\vec\varrho_q = {\mathcal M}_q\vec\rho_q$. The
long range of Coulomb interaction renders the matrix
${\mathcal M}_q$ universal in the limit
$q\to 0$~\cite{wen:prb:94}, and we find as normal modes a)~the
classical~\cite{volkmikh} edge-magnetoplasmon mode,
$\rho^{\text{(emp)}} = \varrho^{\text{(B)}}+\varrho^{\text{(R)}}
+\varrho^{\text{(W)}}$, which is right-moving, and b)~two
linearly dispersing neutral modes, a right-moving one given by
$\rho^{\text{(rn)}}=(\varrho^{\text{(B)}}-\varrho^{\text{(R)}})/
\sqrt{2}$, and the left-moving neutral mode $\rho^{\text{(ln)}}=
(\varrho^{\text{(B)}} + \varrho^{\text{(R)}} + 2
\varrho^{\text{(W)}})/\sqrt{2}$.

In addition to forward and backscattering, another interaction
process exists which has not been noticed previously. (See
Fig.~\ref{scattfig}.) The full effective 1D Hamiltonian
describing edge excitations at a reconstructed edge is actually
given by $H = H_{\text{TL}} + H_{\text{U}}$ with
\widetext
\top{-2.8cm}
\begin{equation}
H_{\text{U}} = \int\int dx\, dx^\prime\,\, V_{\text{U}}(x-
x^\prime)\left\{\big[\psi^{\text{(R)}}(x)\big]^\dagger
\big[\psi^{\text{(B)}}(x^\prime)\big]^\dagger
\psi^{\text{(W)}}(x^\prime) \, \psi^{\text{(W)}}(x) \,\,
e^{i D\frac{x-x^\prime}{2} - i \delta\frac{x+x^\prime}{2}}+
{\mathrm{H.c.}} \right\}\, .
\end{equation}
\bottom{-2.7cm}
\narrowtext
\noindent
Here we have used the abbreviations $D=k_{\text{F}}^{\text{(B)}}
-k_{\text{F}}^{\text{(R)}}$ and $\delta=k_{\text{F}}^{\text{(B)}}
+k_{\text{F}}^{\text{(R)}}-2k_{\text{F}}^{\text{(W)}}$. As the
scattering process represented by $H_{\text{U}}$ converts two
electrons from the left-moving W-branch into electrons belonging
to the right-moving R,B-branches (and {\it vice versa\/}), it is
reminiscent of Umklapp scattering which is important in lattice
models for conventional 1D electron systems near
half-filling~\cite{umlit}. That this analogy reaches quite far
can be seen from the fact that there is a commensuration issue
for the novel Umklapp process at QH edges. It is most relevant if
$k_{\text{F}}^{\text{(B)}}-k_{\text{F}}^{\text{(W)}}=
k_{\text{F}}^{\text{(W)}}-k_{\text{F}}^{\text{(R)}}$, and the
parameter $\delta$ arises naturally as a measure for the
deviation from perfect commensuration~\cite{convumk}. (See
below.)

Umklapp processes do not conserve particle number in each
edge branch separately. Therefore, $H_{\text{U}}$ cannot be
written in terms of a TL model. However, using the bosonization
identity~\cite{bosid,vondelft} for 1D fermionic operators,
\begin{mathletters}\label{singlebosa}
\begin{eqnarray}
\psi^{\text{(R)}}(x) &=&1/ \sqrt{L}\,\,\, {\mathbf :}\,\exp [
i\,\phi^{\text{(R)}}(x)] \, {\mathbf :} \quad ,\\
\psi^{\text{(W)}}(x) &=& 1/\sqrt{L}\,\,\, {\mathbf :}\,\exp [
-i\, \phi^{\text{(W)}}(x)] \, {\mathbf :}\quad ,\\ 
\psi^{\text{(B)}}(x) &=& 1/\sqrt{L}\,\,\, {\mathbf :}\,\exp [
i\, \phi^{\text{(B)}}(x)] \, {\mathbf :}\quad ,
\end{eqnarray}
\end{mathletters}
where ${\mathbf :}\dots{\mathbf :}$ symbolizes normal ordering,
and
\begin{equation}\label{phifields}
\phi^{(\alpha)}(x) = i\, \frac{2\pi}{L} \sum_{q\ne 0}
\frac{e^{- i q x}}{q}\,\, \varrho_q^{(\alpha)}\quad ,
\end{equation}
with $\alpha\in\{\text{R,W,B}\}$, it is possible to rewrite
$H_{\text{U}}$ entirely in terms of bosonic degrees of freedom:
\begin{eqnarray}\label{bosum}
& & H_{\text{U}} = 2 \Lambda^2 g_{\text{U}} \nonumber \\
& & \times \int d x \, \cos\left[
\phi^{\text{(R)}}(x) +\phi^{\text{(B)}}(x) + 2\phi^{\text{(W)}}
(x) + \delta x\right]\, .
\end{eqnarray}
Both the coupling constant $g_{\text{U}}$ and the
incommensuration parameter $\delta$ can be determined from a
microscopic calculation~\cite{uzahm}, and $\Lambda\lesssim (D
\ell^2)^{-1}$ is a physical ultraviolet cut-off. From
Eq.~(\ref{bosum}), it is immediately obvious that $H_{\text{U}}$
couples only to one of the three normal modes of $H_{\text{TL}}$,
namely the left-moving neutral mode $\rho^{\text{(ln)}}$. Hence,
the full Hamiltonian of low-energy excitations at a reconstructed
edge is the sum of three terms, $H=H_{\text{emp}}+H_{\text{rn}}+
H_{\text{ln}}$, where $H_{\text{emp}}$ and $H_{\text{rn}}$
describe the dynamics of free chiral bosons that are associated
with the edge-magnetoplasmon and right-moving neutral modes,
respectively, and
\begin{eqnarray}\label{nphham}
H_{\text{ln}} &=& \int dx \,\left\{ \frac{\hbar v_{\text{ln}}}
{4\pi} \left[\partial_x \,\phi^{\text{(ln)}}(x) \right]^2 \right.
\nonumber \\  & & \hspace{1cm} + 2 \left.
\Lambda^2 g_{\text{U}}\cos\left[\sqrt{2}\,\phi^{\text{(ln)}}(x)
+ \delta x \right]\right\}\, .
\end{eqnarray}
Here, $\phi^{\text{(ln)}}(x)$ is defined in terms of
$\rho^{\text{(ln)}}$ as in Eq.~(\ref{phifields}), and
$v_{\text{ln}}$ is the velocity of the left-moving neutral mode.

Note that Eq.~(\ref{nphham}) does {\em not\/} correspond to the
Hamiltonian of the familiar sine-Gordon model; but rather to a
chiral version of it. To be able to evaluate
electronic correlation functions, we employ a refermionization
technique~\cite{referm} that has been used before to study the
effect of disorder on transport at hierarchical QH
edges~\cite{mpaf:prl:94}. We introduce an auxiliary {\em ghost\/}
field $\eta(x)$ that has the same chirality as the real bosonic
field $\phi^{\text{(ln)}}(x)$ and whose dynamics is given by the
first term of $H_{\text{ln}}$ in Eq.~(\ref{nphham}):
\begin{equation}
H_\eta = \frac{\hbar v_{\text{ln}}}{4\pi} \int dx \,\left[
\partial_x \, \eta(x)\right]^2 \quad .
\end{equation}
The Hamiltonian $H^\prime = H_{\text{ln}} + H_\eta$ is then
equivalent to the bosonized representation of a model of chiral
1D spin-1/2 fermions $\Psi=(\Psi_+ , \Psi_-)$ subject
to a magnetic field $\vec h(x)$ that couples to the
(pseudo-)spin~\cite{pseudospin} degrees of freedom,
\begin{mathletters}
\begin{equation}
H^\prime = \int d x\,\,\,\Psi^\dagger (x) \left\{i\, \hbar\, 
v_{\text{ln}}\, \partial_x \,\, {\mathbf 1} + \vec h(x) \cdot
\vec {\mathbf\sigma} \right\} \Psi(x)\,\, ,
\end{equation}
provided we define $\vec h(x)=\Lambda g_{\text{U}}\,(\cos[\delta
x],\sin[\delta x],0)$ and $\Psi_\pm(x) = 1/\sqrt{L}\,\,\,
{\mathbf :}\,\exp\{i[\eta(x)\pm\phi^{\text{(ln)}}(x)]/\sqrt{2}\}
\,{\mathbf :}$. (We denoted the vector of Pauli matrices by $\vec
{\mathbf\sigma}$.) The introduction of the ghost field $\eta(x)$
turns out to be favorable because it is possible to calculate
physical observables more easily in the refermionized
representation of $H^\prime$ than in the original bosonic theory
with $H_{\text{ln}}$. Note that real physical observables do not
depend on the auxiliary field $\eta(x)$. In Fourier space, the
Hamiltonian $H^\prime$ reads
\begin{equation}
H^\prime = - \hbar\,v_{\text{ln}} \sum_{k s} k\, c^\dagger_{ks}
c_{ks} + \Lambda g_{\text{U}} \sum_k \left[c^\dagger_{k-\delta,+}
c_{k,-} + \mathrm{H.c.}\right]\,\, ,
\end{equation}
and is easily diagonalized, yielding
\begin{equation}
H^\prime = \sum_{ks}\left\{-\hbar\, v_{\text{ln}}\, k + s\,\Delta
/2\right\} \, \varphi_{ks}^\dagger \varphi_{ks} \quad .
\end{equation}
\end{mathletters}
The `Zeeman splitting' induced by the fictitious magnetic field
$\vec h(x)$ is $\Delta = 2 \Lambda g_{\text{U}} [\sqrt{1+\xi^2} -
\xi]$ where $\xi=\hbar v_{\text{ln}}\delta/(2\Lambda g_{\text{U}}
)$ is a measure of the incommensuration. In the ground state, the
free fictitious fermions $\varphi_{ks}$ form two Fermi seas, one
for each spin direction, having different Fermi wave vector due
to the Zeeman splitting.

Having diagonalized $H^\prime$, we are now in the position to
calculate dynamic correlation functions for adding electrons at a
reconstructed edge. We consider the real-time correlation
functions
\begin{equation}
{\mathcal G}^{(\alpha)}(x, t) = \left\langle \psi^{(\alpha)}(x,t)
\,\big[\psi^{(\alpha)}\big]^\dagger(0,0)\right\rangle\quad ,
\end{equation}
where $\alpha\in\{\mbox{R,W,B}\}$. Within the bosonized
representation [Eqs.~(\ref{singlebosa})] of fermionic operators,
the correlation functions ${\mathcal G}^{(\alpha)}(x, t)$
factorize into a product of correlation functions in each of the
three normal modes of $H_{\text{TL}}$:
\begin{mathletters}
\begin{eqnarray}
{\mathcal G}^{(\alpha)}(x, t) &=& \Lambda \,
{\mathcal G}^{(\alpha)}_{\text{emp}}(x, t)\,
{\mathcal G}^{(\alpha)}_{\text{rn}}(x, t)\, 
{\mathcal G}^{(\alpha)}_{\text{ln}}(x, t)\quad , \\
{\mathcal G}^{(\alpha)}_{\beta}(x,t) &=& \left\langle e^{-i
\lambda^{(\alpha)}_{\beta}\phi^{(\beta)}(x,t)} e^{i 
\lambda^{(\alpha)}_{\beta}\phi^{(\beta)}(0,0)}\right\rangle
\quad ,
\end{eqnarray}
\end{mathletters}
with $\beta\in\{\mbox{emp,rn,ln}\}$. The coefficients
$\lambda^{(\alpha)}_{\beta}$ can be read off from the matrix
${\mathcal M}_q$ that relates the density fluctuations at the
R,W,B branches to the normal modes of $H_{\text{TL}}$. As the
edge-magnetoplasmon mode and the right-moving neutral mode are
free bosons, the calculation of 
${\mathcal G}^{(\alpha)}_{\text{emp}}(x,t)$ and
${\mathcal G}^{(\alpha)}_{\text{rn}}(x,t)$ is
standard~\cite{vondelft}, yielding
\begin{equation}\label{powerlaw}
{\mathcal G}^{(\alpha)}_{\beta}(x,t) = \left[\Lambda (x - t
v_{\beta})\right]^{-[\lambda^{(\alpha)}_{\beta}]^2}
\end{equation}
with $\beta\in\{\text{emp,rn}\}$. However, the left-moving
neutral mode is {\em not\/} free due to Umklapp scattering, and
the calculation of ${\mathcal G}^{(\alpha)}_{\text{ln}}(x,t)$ is
nontrivial. Now our refermionized representation turns out to be
useful because we can calculate
${\mathcal G}^{\text{(W)}}_{\text{ln}}(x,t)$ {\em exactly\/}
using the identity
\begin{equation}\label{usereferm}
{\mathcal G}^{\text{(W)}}_{\text{ln}}(x,t) = \Lambda^{-2}\left
\langle\Psi_+^\dagger(x,t)\Psi_-(x,t)\Psi_-^\dagger(0,0)
\Psi_+(0,0)\right\rangle\,\, .
\end{equation}
An elementary calculation in the representation of the free
fermions $\varphi_{k s}$ yields the particle-hole correlation
function for fictitious fermions shown on the r.h.s.\ of
Eq.~(\ref{usereferm}). The leading term in the long-time,
large-distance limit turns out to be a constant,
\begin{mathletters}
\begin{eqnarray}
{\mathcal G}^{\text{(W)}}_{\text{ln}}(x,t) &=& C_{\text{ln}}^2 +
{\mathcal O}\big([x+ t v_{\text{ln}}]^{-2}\big) \quad , \\
C_{\text{ln}} &=&\frac{g_{\text{U}}}{2\pi \hbar v_{\text{ln}}}
\left( 1 - \frac{\xi}{\sqrt{1+\xi^2}}\right)\quad .
\end{eqnarray}
\end{mathletters}
In the limit of large deviation from commensuration, i.e,
$\xi\to\infty$, the constant $C_{\text{ln}}$ vanishes and the
leading behavior of ${\mathcal G}^{\text{(W)}}_{\text{ln}}(x,t)$
is given by the standard result in the absence of Umklapp
scattering.

Unlike for ${\mathcal G}^{\text{(W)}}_{\text{ln}}(x,t)$, there is
no simple representation of
${\mathcal G}^{\text{(R)}}_{\text{ln}}(x,t)$ and
${\mathcal G}^{\text{(B)}}_{\text{ln}}(x,t)$ in terms of
fictitious fermions. This spoils the possibility to evaluate
these correlation functions exactly. However, applying arguments
that are familiar~\cite{gut:prb:76} from the study of
conventional 1D electron systems, we conjecture
\begin{equation}
{\mathcal G}^{\text{(B)}}_{\text{ln}}(x,t) \approx
{\mathcal G}^{\text{(R)}}_{\text{ln}}(x,t) \approx C_{\text{ln}}
\quad .
\end{equation}

The tunneling density of states ${\mathcal A}^{(\alpha)}
(\varepsilon)$ for adding electrons at the branch $\alpha\in\{
\text{R,W,B}\}$ can be calculated straightforwardly~\cite{mahan}
from ${\mathcal G}^{(\alpha)}(0, t)$. We find
\begin{equation}
{\mathcal A}^{(\alpha)}(\varepsilon)\propto \left\{
\begin{array}{cl}
\varepsilon^{[\lambda^{(\alpha)}_{\text{emp}}]^2 + [\lambda^{(
\alpha)}_{\text{rn}}]^2-1} & \mbox{for } \varepsilon < \Delta \\
\varepsilon^{[\lambda^{(\alpha)}_{\text{emp}}]^2 +
[\lambda^{(\alpha)}_{\text{rn}}]^2 + [\lambda^{(\alpha
)}_{\text{ln}}]^2-1} & \mbox{for }\varepsilon > \Delta\end{array}
\right. \, .
\end{equation}
We see that Umklapp scattering at reconstructed QH edges leads to
a {\em suppression\/} of Luttinger-liquid behavior at energies
smaller than $\Delta$. This suggests a scenario for the
experimental verification of Umklapp scattering. As it is the
long range of Coulomb interaction that makes Umklapp scattering
possible in the first place, screening by a nearby metallic gate
will suppress it. The clearest indication for the presence of
Umklapp scattering would be gained in the measurement of the
edge-tunneling exponent~\cite{amc:prl} for various distances
$\lambda$ of the gate. The exponent should have a
non-monotonic dependence on $\lambda$, showing a peak for
intermediate distances $\ell <\lambda < D$. In contrast, the
exponent would be a monotonously increasing function of $\lambda$
in the absence of Umklapp scattering.

In conclusion, we have studied low-energy excitations at a
reconstructed QH edge and identified an Umklapp process that
has not been discussed previously. We solved the theory including
Umklapp exactly and evaluated electronic correlation functions.
It turns out that Umklapp scattering suppresses Luttinger-liquid
behavior.

This work was funded in part by NSF Grant No.\ DMR-9714055 and
Sonderforschungsbereich 195 der Deutschen Forschungsgemeinschaft.
I would like to thank A.~H.~MacDonald for suggesting the
Umklapp-scattering project and for his constant help and
encouragement. I benefited from discussions with I.~Affleck,
W.~Apel, C.~de~C.~Chamon, E.~Fradkin, S.~M.~Girvin,
B.~I.~Halperin, V.~Meden, N.~P.~Sandler, and J.~von~Delft.

{\em Note added:\/} After submission of this work, I became aware
of an independent recent study of chiral sine-Gordon
theory~\cite{sls:99}.

\widetext

\begin{thebibliography}{10}

\bibitem[*]{preadd}
Present address.

\bibitem{volkmikh}
V.~A. Volkov and S.~A. Mikhailov,  in {\em Landau Level
Spectroscopy}, edited by G. Landwehr and E.~I. Rashba (Elsevier
Science, Amsterdam, 1991), pp.\  855--907.

\bibitem{ahmintro}
A.~H. MacDonald,  in {\em Mesoscopic Quantum Physics},
{\em Proceedings of the 1994 Les Houches Summer School, Session
LXI}, edited by E. Akkermans {\it  et~al.} (Elsevier Science,
Amsterdam, 1995), pp.\ 659--720.

\bibitem{sharpness}
The sharpness of the edge can be quantified by the slope of the
bare confining potential at the sample boundary.

\bibitem{ahm:wen:90}
A.~H. MacDonald, Phys. Rev. Lett. {\bf 64},  220  (1990);
X.~G. Wen, Phys. Rev. B {\bf 41},  12838  (1990).

\bibitem{wen:rev}
X.~G. Wen, Int. J. Mod. Phys. B {\bf 6},  1711  (1992).

\bibitem{tom:lutt}
S. Tomonaga, Prog. Theor. Phys. {\bf 5},  544  (1950);
J.~M. Luttinger, J. Math. Phys. {\bf 4},  1154  (1963).

\bibitem{rightmove}
Specifying the chirality of edge excitations fixes the direction
of the perpendicular magnetic field.

\bibitem{ahm:aust:93}
A.~H. MacDonald, S.~R. Yang, and M.~D. Johnson, Aust. J. Phys.
{\bf 46},  345  (1993).

\bibitem{wen:prb:94}
C. de~C.~Chamon and X.~G. Wen, Phys. Rev. B {\bf 49}, 8227
(1994).

\bibitem{otherrec}
For related work on edge reconstruction in spin-polarized QH
samples, see, e.g., Y. Meir, Phys. Rev. Lett. {\bf 72}, 2624
(1994); L. Brey, Phys. Rev. B  {\bf 50}, 11861 (1994); D.~B.
Chklovskii, Phys. Rev. B {\bf 51}, 9895 (1995).  Scenarios for
edge reconstruction involving the spin-degree of freedom have
been proposed as well. [See, e.g., J. Dempsey, B.~Y. Gelfand, and
B.~I. Halperin, Phys. Rev. Lett. {\bf 70}, 3639 (1993); A.~S.
Karlhede {\it et~al.}, Phys. Rev. Lett. {\bf 77}, 2061 (1996);
J.~H. Oaknin, L. Martin-Moreno, and C. Tejedor, Phys. Rev. B
{\bf 54}, 16850 (1996).] Here we assume that electrons occupy
only states in the lowest spin-split Landau level and neglect
spin altogether.

\bibitem{exprec}
Recent experimental studies addressing edge reconstruction are,
e.g., O. Klein {\it et~al.}, Phys. Rev. Lett. {\bf 74}, 785
(1995); Phys. Rev. B {\bf 53}, R4221 (1996); R.~C. Ashoori,
Nature (London) {\bf 379}, 413 (1996); T.~H. Osterkamp
{\it et~al.}, Phys. Rev. Lett. {\bf 82}, 2931 (1999).

\bibitem{rennapology}
Recent numerical studies of edge reconstruction in quantum dots
[S.~M. Reimann {\it et al.}, Phys. Rev. Lett. {\bf 83}, 3270
(1999); E. Goldmann and S.~R. Renn, cond-mat/9909071] find a
tendency toward Wigner crystallization along the edge. At
present, it is not clear whether this tendency persists for
bulk systems and how it depends on details of the confining
potential. Here we consider the case of uniformly reconstructed
edges only.

\bibitem{manymodes}
I.~L. Aleiner and L.~I. Glazman, Phys. Rev. Lett. {\bf 72},  2935
(1994); S. Conti and G. Vignale, Phys. Rev. B {\bf 54},  R14309
(1996); J.~H. Han and D.~J. Thouless, Phys. Rev. B {\bf 55},
R1926  (1997).

\bibitem{smooth}
See, e.g., C.~W.~J. Beenakker, Phys. Rev. Lett. {\bf 64}, 216
(1990); A.~M. Chang, Solid State Commun. {\bf 74}, 871 (1990);
D.~B. Chklovskii, B.~I. Shklovskii, and L.~I. Glazman, Phys. Rev.
B {\bf 46}, 4026 (1992); K. Lier and R.~R. Gerhardts, Phys. Rev.
B {\bf 50}, 7757 (1994).

\bibitem{mpaf:prb:95a}
C.~L. Kane and M.~P.~A. Fisher, Phys. Rev. B {\bf 51},  13449
(1995).

\bibitem{amc:prl}
A.~M. Chang, L.~N. Pfeiffer, and K.~W. West, Phys. Rev. Lett.
{\bf 77},  2538 (1996); M. Grayson {\it et~al.}, Phys. Rev. Lett.
{\bf 80},  1062  (1998).

\bibitem{sol:adv:79}
J. S\a'olyom, Adv. Phys. {\bf 28},  201  (1979).

\bibitem{backscatt}
For spinless electrons, backscattering simply renormalizes
forward-scattering amplitudes.

\bibitem{umlit}
V.~J. Emery,  in {\em Highly Conducting One-Dimensional Solids},
edited by J.~T. Devreese {\it et~al.} (Plenum Press, New York,
1979), pp.\ 247--303; R. Shankar, Int. J. Mod. Phys. B {\bf 4},
2371 (1990); H.~J. Schulz, in {\em Strongly Correlated Electronic
Materials}, edited by K.~S. Bedell {\it et~al.} (Addison Wesley,
Reading, MA, 1994), pp.\ 187--232.

\bibitem{convumk}
In conventional 1D electron systems, Umklapp scattering
necessarily involves a large momentum transfer $\sim 4
k_{\text{F}}$ that can be absorbed by a reciprocal lattice
vector only near half-filling. This is different for the novel
Umklapp process at reconstructed QH edges where momentum
transfers can be small and no underlying lattice exists.

\bibitem{bosid}
F.~D.~M. Haldane, J. Phys. C {\bf 14},  2585  (1981);
J. Voit, Rep. Prog. Phys. {\bf 57},  977  (1994);
R. Shankar, Acta Phys. Pol. B {\bf 26},  1835  (1995).

\bibitem{vondelft}
J. {von Delft} and H. Schoeller, Ann. Phys. (Leipzig) {\bf 7}, 
225 (1998).

\bibitem{uzahm}
U. Z\"ulicke and A.~H.~MacDonald, Physica E (to appear).

\bibitem{referm}
The method of refermionization was pioneered by Luther and Emery
[Phys. Rev. Lett. {\bf 33}, 589 (1974)]. For a review and recent
applications, see Ref.~\onlinecite{vondelft}.

\bibitem{mpaf:prl:94}
C.~L. Kane, M.~P.~A. Fisher, and J. Polchinski, Phys. Rev. Lett.
{\bf 72}, 4129  (1994).

\bibitem{pseudospin}
The spin degree of freedom of the fictitious fermions is
unrelated to the spin of electrons in the QH sample.

\bibitem{gut:prb:76}
H. Gutfreund and R.~A. Klemm, Phys. Rev. B {\bf 14}, 1073 (1976).

\bibitem{mahan}
G.~D. Mahan, {\em Many-Particle Physics} (Plenum Press, New York,
1990).

\bibitem{sls:99}
J.~D. Naud, L.~P. Pryadko, and S.~L. Sondhi, cond-mat/9908188.

\end{thebibliography}
\end{document}